# Inducing Kondo Screening of Vacancy Magnetic Moments in Graphene with Gating and Local Curvature

Yuhang Jiang[1], Po-Wei Lo[2,3,#], Daniel May[4], Guohong Li[1], Guang-Yu Guo[2,3], Frithjof B. Anders[4], Takashi Taniguchi[5], Kenji Watanabe[5], Jinhai Mao[1,6] and Eva Y. Andrei[1]

[1]Department of Physics and Astronomy, Rutgers University, 136 Frelinghuysen Road, Piscataway, NJ 08855 USA

[2]Department of Physics, National Taiwan University, Taipei 10617, Taiwan

[3]Physics Division, National Center for Theoretical Sciences, Hsinchu 30013, Taiwan

[4]Theoretische Physik 2, Technische Universität Dortmund, 44221 Dortmund, Germany

[5]Advanced Materials Laboratory, National Institute for Materials Science, 1-1 Namiki, Tsukuba 305-0044, Japan

[6]Institute of Physics & University of Chinese Academy of Sciences, Beijing 100190, China

Abstract

In normal metals the magnetic moment of impurity-spins disappears below a characteristic Kondo temperature which marks the formation of a cloud of conduction-band electrons that screen the local-moment. In contrast, moments embedded in insulators remain unscreened at all temperatures. What then is the fate of magnetic-moments in intermediate, pseudogap systems, such as graphene? Theory predicts that coupling to the conduction-band electrons will drive a quantum phase transition between a local-moment phase and a Kondo-screened phase. However, attempts to experimentally confirm this prediction and its intriguing consequences, such as electrostatically tunable magnetic-moments, have been elusive. Here we report the observation of Kondo-screening and the quantum phase-transition between screened and unscreened phases of vacancy magnetic moments in graphene. Using scanning tunneling spectroscopy and numerical renormalization-group calculations we show that this transition enables to control the screening of local moments by tuning the gate voltage and the local curvature of the graphene membrane.

## Introduction

Graphene, with its linear DOS and tunable chemical potential [1,2], provides a playground for exploring the physics of the magnetic quantum phase transition [3,4,5,6,7,8,9] (Fig. 1a). But embedding a magnetic moment and producing sufficiently large coupling with the itinerant electrons in graphene, poses significant experimental challenges: adatoms typically reside far above the graphene plane, while substitutional atoms tend to become delocalized and non-magnetic [10]. An alternative and efficient way to embed a magnetic moment in graphene is to create single atom vacancies. The removal of a carbon atom from the honeycomb lattice induces a magnetic moment stemming from the unpaired electrons at the vacancy site [11,12,13,14]. This moment has two contributions: one is a resonant state (zero mode - ZM) at the Dirac point (DP) due to the unpaired electron left by the removal of an electron from the π-band; the other arises from the broken σ-orbitals, two of which hybridize leaving a dangling bond that hosts an unpaired electron [14]. The ZM couples ferromagnetically to the dangling σ-orbital [14] as well as to the conduction electrons [15,16] and remains unscreened. In flat graphene the magnetic moment from the dangling σ-bond is similarly unscreened because the σ-orbital is orthogonal to the π-band conduction electrons [16,17]. However, it has been proposed that this constraint would be eliminated in the presence of a local curvature which removes the orthogonality of the σ-orbital with the conduction band, and enables Kondo screening [16,18,19]. One strategy to introduce local curvature is to deposit the flexible graphene membrane on a corrugated substrate.

Here we employ the spectroscopic signature of the Kondo effect to demonstrate that screening of vacancy magnetic moments in graphene is enabled by corrugated substrates. Crucially, variations in the local curvature imposed by the corrugated substrate provide a range of coupling strengths, from subcritical to the supercritical regime, all in the same sample. An unexpected consequence of this unusually wide variation of coupling strengths is that global measurements such as magnetization [20,21] or resistivity [22] can give contradictory results. In fact, as we show, the quantum critical transition between Kondo screened and local moment phases in this system, can only be observed through a local measurement.

## Results

### Scanning tunneling microscopy and spectroscopy

We employed scanning tunneling spectroscopy (STS) [23, 24] to identify Kondo screening of the vacancy magnetic moment by the distinctive zero-bias resonance it produces in the *dI/dV* curves (*I* is the tunneling current and *V* the junction bias), hereafter called Kondo peak. We first discuss samples consisting of two stacked single layer graphene sheets on a $SiO_2$ substrate (G/G/$SiO_2$) capping a doped Si gate electrode (Fig. 1b). A large twist angle between the two layers ensures electronic decoupling, and preserves the electronic structure of single layer graphene while reducing substrate induced random potential fluctuations [24, 25, 26]. A further check of the Landau-level spectra in a magnetic field revealed the characteristic sequence expected for massless Dirac fermions [2, 23], confirming the electronic decoupling of the two layers (Supplementary Note 1). Vacancies were created by low energy (100 eV) $He^+$ ion sputtering followed by in situ annealing [22, 27, 28]. In STM topography of a typical irradiated sample (Fig. 1c) the vacancies appear as small protrusions on top of large background corrugations. To establish the nature of a vacancy we zoom in to obtain atomic resolution topography and spectroscopy. Single atom vacancies are recognized by their distinctive triangular $\sqrt{3} \times \sqrt{3} \, R30°$ topographic fingerprint (Fig. 1c inset) [27, 28, 29] which is accompanied by a pronounced peak in the *dI/dV* spectra at the DP reflecting the presence of the ZM. If both these features are present we identify the vacancy as a single atom vacancy (Supplementary Note 2) and proceed to study it further. In order to separate the physics at the DP and the Kondo screening which produces a peak near Fermi energy, $E_F \equiv 0$, the spectrum of the vacancy in Fig. 1d is taken at finite doping corresponding to a chemical potential, $\mu \equiv E_F - E_D = -54 \, meV$. Far from the vacancy (lower curve), we observe the V shaped spectrum characteristic of pristine graphene, with the minimum identifying the DP energy. In contrast, at the center of the vacancy (Fig. 1d upper curve), the spectrum features two peaks, one at the DP identifying the ZM and the other at zero bias coincides with the position of the expected Kondo peak [3]. (In STS the zero-bias is identified with $E_F$.) From the line shape of the zero-bias peak (Fig. 1f inset), we extract $T_K = (67 \pm 2)$ K by fitting to the Fano line shape [30, 31] characteristic of Kondo resonances (Supplementary Note 3). As a further independent check we compare in Fig. 1f the temperature dependence of the linewidth to that expected for a Kondo-screened impurity [30, 32] (Supplementary

Note 4), $\Gamma_{LW} = \sqrt{(\alpha k_B T)^2 + (2k_B T_K)^2}$ from which we obtain $T_K = (68 \pm 2)$K, consistent with the above value, and $\alpha = 6.0 \pm 0.3$ in agreement with measurements and numerical simulations on ad-atoms [30, 33]. Importantly, as we show below, this resonance is pinned to $E_F$ over the entire range of chemical potential values studied, as expected for the Kondo peak [34, 35].

The gate dependence of the spectra corresponding to the hundreds of vacancies studied here falls into two clearly defined categories, which we label type I, and type II. In Fig. 2a we show the evolution with chemical potential of the spectra at the center of a type I vacancy. Deep in the p-doped regime, we observe a peak which is tightly pinned to, $E_F$, consistent with Kondo-screening. Upon approaching charge neutrality the Kondo peak disappears for $\mu \geq -58$ meV and reenters asymmetrically in the n-doped sector, for $\mu \geq 10$ meV. As we discuss below, the absence of screening close to the charge neutrality point and its reentrance in the n-doped regime for type I vacancies is indicative of pseudogap Kondo physics for subcritical coupling strengths [8, 36]. For type II vacancies, the evolution of the spectra with chemical potential, shown in Fig. 2b, is qualitatively different. The Kondo peak is observed in the p-doped regime and disappears close to charge neutrality, but does not reappear on the n-doped side. We show below that this behavior is characteristic of pseudogap Kondo physics for vacancies whose coupling to the conduction band is supercritical [8, 36].

### Numerical renormalization group calculations

To better understand the experimental results we performed numerical-renormalization-group (NRG) calculations for a minimal model based on the pseudogap asymmetric Anderson impurity model (AIM) [6, 16, 37] comprising the free local σ-orbital coupled to the itinerant π-band (Supplementary Note 6). This single orbital model gives an accurate description of the experiment in the p-doped regime where the ZM is sufficiently far from the Kondo peak so that their overlap is negligible. Upon approaching charge neutrality, interactions between the two orbitals through Hund's coupling and level repulsion become relevant. As described in Supplementary Note 6 we introduced an effective Coulomb interaction term to take into account this additional repulsion. The single orbital model together with this phenomenological correction captures the main features of the Kondo physics reported here (Fig. 3). Results from a comprehensive NRG calculation using a two-orbital pseudogap AIM to model the problem [38] similarly indicate that this simplified one-orbital approach qualitatively describes the experimental results. The single orbital AIM is characterized by three energy scales, $\varepsilon_d$, $U$ and $\Gamma_0$, corresponding to the energy of the impurity state, the onsite Coulomb repulsion, and by the scattering rate or exchange between the impurity and the conduction electrons, respectively (Supplementary Note 7). In the asymmetric AIM, which is relevant to screening of vacancy magnetic moments in graphene, the particle-hole symmetry is broken by next-nearest neighbor hopping and by $U \neq 2|\varepsilon|$. The NRG phase diagram for this model is controlled by the valence fluctuation (VF) critical point, $\Gamma_C$ [6, 7, 8, 39, 40]. At charge neutrality ($\mu = 0$), $\Gamma_C$ separates the NRG flow into two sectors: supercritical, $\Gamma_0 > \Gamma_C$, which flows to the asymmetric strong-coupling (ASC) fixed point where charge fluctuations give rise to a frozen impurity (FI) ground state [41], and subcritical, $\Gamma_0 < \Gamma_C$, which flows to the local moment (LM) fixed point where the impurity moment is unscreened. At the FI fixed point, the correlated ground state acquires one additional charge due to the enhancement of the particle-hole asymmetry in the RG flow. In a simplified picture, the fixed point spectrum can be understood by the flow of $\varepsilon_d \to -\infty$, leading to effective doubly occupied singlet impurity state that decouples from the remaining conduction band [6, 8, 41]. In terms of the real physical orbitals, however, the NRG reveals a distribution of this additional charge between the conduction band and the local orbital with a small enhancement of $n_\sigma = 1.2$-$1.3$. For $\Gamma_0 < \Gamma_C$ and $\mu \neq 0$, the appearance of relevant spin fluctuations gives rise to a cloud of spin-polarized electrons that screen the local moment below a characteristic temperature $T_K$ which is exponentially suppressed[8] ($\ln T_K \propto -1/|\mu|$). As a result, at sufficiently low doping, $T_K$ must fall below any experimentally accessible temperature, so that for all practical purposes its value can be set to zero (Fig. 4a). Using NRG to simulate the experimental spectra (Supplementary Note 7) we found $\varepsilon_d = -1.6$ eV for the bare σ-orbital energy [11, 36], $U = 2$ eV [11, 42, 43] and a critical coupling $\Gamma_C = 1.15$ eV that separates the LM and the FI phases at $\mu = 0$. From the NRG fits of the STS spectra we obtained the value of the reduced coupling $\Gamma_0/\Gamma_C$ for each vacancy shown in Fig. 3 (Supplementary Note 8). The values, $\Gamma_0/\Gamma_C = 0.90$, and 1.83 obtained for the spectra in Fig. 2a and 2b place these two vacancies in the sub-critical and super-critical regimes respectively.

In Fig. 3 we compare the chemical-potential dependence of the measured $T_K$, with the NRG results. The $T_K$ values are obtained from Fano-fits of the Kondo peaks leading to the $T_K(\mu)$ curves, shown in Fig. 3a. The corresponding values of $\Gamma_0/\Gamma_C$ and the $T_K(\mu)$ curves obtained by using NRG to simulate the spectra are shown in Fig. 3b. The close agreement between experiment and simulations confirms the validity of the asymmetric AIM for describing screening of vacancy spins in graphene. In Fig. 4a we summarize the numerical results in a $\mu$-$\Gamma_0$ phase diagram. At charge neutrality (defined by the $\mu = 0$ line), the critical point $\Gamma_0/\Gamma_C = 1$ signals a quantum phase transition between the LM phase and the FI phase [36]. The Kondo-screened phase appears at finite doping ($\mu \neq 0$) and is marked by the appearance of the Kondo-peak, [8, 9]. The phase diagram clearly shows the strong electron-hole asymmetry consistent with the asymmetric screening expected in in this system [4].

## Dependence of Kondo screening on corrugation amplitude

Theoretical work [16, 44] suggests that coupling of the vacancy moment with the conduction electrons in graphene may occur if local corrugations produce an out of plane component of the dangling σ-orbital. This removes the orthogonality restriction[17] that prevents hybridization of the σ-bands and π-bands in flat graphene and produces a finite coupling strength which increases monotonically with the out of plane projection of the orbital[18, 19, 45]. To check this conjecture we repeated the experiments for samples on substrates with different average corrugation amplitudes as shown in Fig. 4b and 4c. For consistency all the fabrication steps were identical. In the G/SiO$_2$ sample (single layer graphene on SiO$_2$) where the corrugation amplitude was largest (~1nm), 60% of the vacancies displayed the Kondo peak and $T_K$ attained values as high as 180K (Supplementary Note 5). For the flatter G/G/SiO$_2$ where the average corrugation was ~0.5nm, we found that 30% of the vacancies showed the Kondo peak with $T_K$ values up to 70K. For samples deposited on hBN, which were the flattest with local corrugation amplitudes of ~0.1nm, none of the vacancies showed the Kondo peak. This is illustrated in Fig. 1e showing a typical $dI/dV$ curve on a vacancy in G/G/BN (double layer graphene on hBN) where a gate voltage of $V_g = -30$V was applied to separate the energies of $E_F$ and the DP. While this spectrum shows a clear ZM peak, the Kondo peak is absent over the entire range of doping[28]. The absence of the Kondo peak in all the samples deposited on hBN highlights the importance of the local curvature. In order to quantify the effect of the local curvature on the coupling strength, we employed STM topography to measure the local radius of curvature, $R$, at the vacancy sites (Fig. 4d inset) from which we estimate the angle between the σ-orbital and the local graphene plane orientation [45], $\theta \approx a/2R$, where $a$ is the lattice spacing. We find that the coupling strength, $\Gamma_0/\Gamma_C(\theta)$, shows a monotonic increase with $\theta$ (Fig.

4d), consistent with the theoretical expectations [16, 18, 45]. Interestingly, the effect of the curvature on the Kondo coupling was also observed for Co atoms deposited on corrugated graphene [46], and was also utilized to enhance the spin-orbit coupling [47].

## Discussion

The results presented here shed light on the contradictory conclusions drawn from earlier magnetometry [20, 21] and transport measurements [22] on irradiation induced vacancies in graphene. While the transport measurements revealed a resistivity minimum and logarithmic scaling indicative of Kondo screening with unusually large values of $T_K \sim 90K$, magnetometry measurements showed Curie behavior with no evidence of low-temperature saturation, suggesting that the vacancy moments remained unscreened. To understand the origin of this discrepancy we note that magnetometry and transport are sensitive to complementary aspects of the Kondo effect. The former probes the magnetic moment and therefore only sees vacancies that are not screened, while the latter probes the enhanced scattering from the Kondo cloud which selects only the vacancies whose moment is Kondo screened. Importantly, these techniques take a global average over all the vacancies in the sample. This does not pose a problem when all the impurities have identical coupling strengths. But if there is a distribution of couplings ranging from zero to finite values, as is the case here, global magnetization and transport measurements will necessarily lead to opposite conclusions as reported in the earlier work.

The local spectroscopy technique employed here made it possible to disentangle the physics of Kondo screening in the presence of a distribution of coupling strengths. This work demonstrates the existence of Kondo screening in a pseudogap system and identifies the quantum phase transition between a screened and an unscreened local magnetic moment. It further shows that the local magnetic moment can be tuned both electrically and mechanically, by using a gate voltage and a local curvature respectively.

## Methods

### Sample Fabrication

The G/G/SiO$_2$ samples consisted of two stacked graphene layers deposited on a 300nm SiO$_2$ dielectric layer capping a highly doped Si chip (acting as the backgate electrode)[23, 48, 49]. The bottom graphene layer was exfoliated onto the SiO$_2$ surface and the second layer was stacked on top by a dry transfer process. PMMA and PVA thin films were used as the carrier in the dry transfer process. Au/Ti electrodes were added by standard SEM lithography, followed by a metal thermal deposition process. After liftoff, the sample was annealed in forming gas (H$_2$ : Ar, 1:9) at 300℃ for 3 hours to remove the PMMA residue, and further annealed overnight at 230 ℃ in UHV[23]. All other samples (G/SiO$_2$, G/G/BN and G/BN) were fabricated by a similar layer-by-layer dry transfer process. To introduce single vacancies in the graphene lattice, the device was exposed under UHV conditions to a beam of He$^+$ ions with energy 100eV for 5 to 10 seconds, and further annealed at high temperature in situ [28].

Scanning Tunneling Microscopy Experiment

Except where mentioned all the STM experiments were performed at 4.2K. dI/dV curves were collected by the standard lock-in technique, with 0.5mV AC modulation at 473Hz added to the DC sample bias[2, 24, 50]. The chemical potential was tuned by the backgate voltage as illustrated in Fig.1b.

References:

1. Castro Neto AH, Guinea F, Peres NMR, Novoselov KS, Geim AK. The electronic properties of graphene. *Rev Mod Phys* **81**, 109-162 (2009).

2. Andrei EY, Li GH, Du X. Electronic properties of graphene: a perspective from scanning tunneling microscopy and magnetotransport. *Rep Prog Phys* **75**, 056501 (2012).

3. Hewson AC. The Kondo Problem to Heavy Fermions. *Cambridge University Press, Cambridge*, (1997).

4. Withoff D, Fradkin E. Phase Transitions in Gapless Fermi Systems with Magnetic Impurities. *Phys Rev Lett* **64**, 1835-1838 (1990).

5. Cassanello CR, Fradkin E. Kondo effect in flux phases. *Phys Rev B* **53**, 15079-15094 (1996).

6. Gonzalez-Buxton C, Ingersent K. Renormalization-group study of Anderson and Kondo impurities in gapless Fermi systems. *Phys Rev B* **57**, 14254-14293 (1998).

7. Fritz L, Vojta M. Phase transitions in the pseudogap Anderson and Kondo models: Critical dimensions, renormalization group, and local-moment criticality. *Phys Rev B* **70**, 214427 (2004).

8. Vojta M, Fritz L, Bulla R. Gate-controlled Kondo screening in graphene: Quantum criticality and electron-hole asymmetry. *Europhysics Letters* **90**, 27006 (2010).

9. Kanao T, Matsuura H, Ogata M. Theory of Defect-Induced Kondo Effect in Graphene: Numerical Renormalization Group Study. *J Phys Soc Jpn* **81**, 063709 (2012).

10. Wang H*, et al.* Doping Monolayer Graphene with Single Atom Substitutions. *Nano Letters* **12**, 141-144 (2012).

11. Yazyev OV, Helm L. Defect-induced magnetism in graphene. *Phys Rev B* **75**, 125408 (2007).

12. Uchoa B, Kotov VN, Peres NMR, Castro Neto AH. Localized magnetic states in graphene. *Phys Rev Lett* **101**, 026805 (2008).

13. Palacios JJ, Fernandez-Rossier J, Brey L. Vacancy-induced magnetism in graphene and graphene ribbons. *Phys Rev B* **77**, 195428 (2008).

14. Nanda BRK, Sherafati M, Popović ZS, Satpathy S. Electronic structure of the substitutional vacancy in graphene: density-functional and Green's function studies. *New J Phys* **14**, 083004 (2012).

15. Haase P, Fuchs S, Pruschke T, Ochoa H, Guinea F. Magnetic moments and Kondo effect near vacancies and resonant scatterers in graphene. *Phys Rev B* **83**, 241408 (2011).

16. M. A. Cazalilla, A. Iucci, F. Guinea, Castro-Neto AH. Local Moment Formation and Kondo Effect in Defective Graphene. Preprint at https://arxiv.org/abs/1207.3135 (2012).

17. Hentschel M, Guinea F. Orthogonality catastrophe and Kondo effect in graphene. *Phys Rev B* **76**, 115407 (2007).

18. Ando T. Spin-orbit interaction in carbon nanotubes. *J Phys Soc Jpn* **69**, 1757-1763 (2000).

19. Castro Neto AH, Guinea F. Impurity-Induced Spin-Orbit Coupling in Graphene. *Phys Rev Lett* **103**, 026804 (2009).

20. Nair RR*, et al.* Dual origin of defect magnetism in graphene and its reversible switching by molecular doping. *Nat Commun*, 4:2010 doi: 2010.1038/ncomms3010 (2013).

21. Nair RR*, et al.* Spin-half paramagnetism in graphene induced by point defects. *Nat Phys* **8**, 199-202 (2012).

22. Chen JH, Li L, Cullen WG, Williams ED, Fuhrer MS. Tunable Kondo effect in graphene with defects. *Nat Phys* **7**, 535-538 (2011).

23. Luican A, Li G, Andrei EY. Scanning tunneling microscopy and spectroscopy of graphene layers on graphite. *Solid State Communications* **149**, 1151-1156 (2009).

24. Li G, Luican A, Andrei EY. Self-navigation of a scanning tunneling microscope tip toward a micron-sized graphene sample. *Review of Scientific Instruments* **82**, 073701 (2011).

25. Li G*, et al.* Observation of Van Hove singularities in twisted graphene layers. *Nat Phys* **6**, 109-113 (2010).

26. Luican A*, et al.* Single-Layer Behavior and Its Breakdown in Twisted Graphene Layers. *Phys Rev Lett* **106**, 126802 (2011).

27. Ugeda MM, Brihuega I, Guinea F, Gómez-Rodríguez JM. Missing Atom as a Source of Carbon Magnetism. *Phys Rev Lett* **104**, 096804 (2010).

28. Mao J*, et al.* Realization of a Tunable Artificial Atom at a Charged Vacancy in Graphene *Nat Phys* **12**, 545-549 (2016).

29. Kelly KF, Sarkar D, Hale GD, Oldenburg SJ, Halas NJ. Threefold Electron Scattering on Graphite Observed with C60-Adsorbed STM Tips. *Science* **273**, 1371-1373 (1996).

30. Ternes M, Heinrich AJ, Schneider WD. Spectroscopic manifestations of the Kondo effect on single adatoms. *Journal of Physics: Condensed Matter* **21**, 053001 (2009).

31. Schiller A, Hershfield S. Theory of scanning tunneling spectroscopy of a magnetic adatom on a metallic surface. *Phys Rev B* **61**, 9036-9046 (2000).

32. Nagaoka K, Jamneala T, Grobis M, Crommie MF. Temperature dependence of a single Kondo impurity. *Phys Rev Lett* **88**, 077205 (2002).

33. Otte AF*, et al.* The role of magnetic anisotropy in the Kondo effect. *Nat Phys* **4**, 847-850 (2008).

34. Cronenwett SM, Oosterkamp TH, Kouwenhoven LP. A tunable Kondo effect in quantum dots. *Science* **281**, 540-544 (1998).

35. Goldhaber-Gordon D, Shtrikman H, Mahalu D, Abusch-Magder D, Meirav U, Kastner MA. Kondo effect in a single-electron transistor. *Nature* **391**, 156-159 (1998).

36. Lo PW, Guo GY, Anders FB. Gate-tunable Kondo resistivity and dephasing rate in graphene studied by numerical renormalization group calculations. *Phys Rev B* **89**, 195424 (2014).

37. Ruiz-Tijerina DA, Dias da Silva LGGV. Transport signatures of Kondo physics and quantum criticality in graphene with magnetic impurities. *Phys Rev B* **95**, 115408 (2017).

38. Daniel May P-WL, Kira Deltenre, Anika Henke, Jinhai Mao, Yuhang Jiang, Guohong Li, Eva Y. Andrei, Guang-Yu Guo, Frithjof B. Anders. Modeling of gate-controlled Kondo effect at carbon point defects in graphene. *Phys Rev B* **97**, 155419 (2018).

39. Mitchell AK, Vojta M, Bulla R, Fritz L. Quantum phase transitions and thermodynamics of the power-law Kondo model. *Phys Rev B* **88**, 195119 (2013).

40. Fritz L, Vojta M. The physics of Kondo impurities in graphene. *Rep Prog Phys* **76**, 032501 (2013).

41. Krishna-murthy HR, Wilkins JW, Wilson KG. Renormalization-group approach to the Anderson model of dilute magnetic alloys. II. Static properties for the asymmetric case. *Phys Rev B* **21**, 1044-1083 (1980).

42. Padmanabhan H, Nanda BRK. Intertwined lattice deformation and magnetism in monovacancy graphene. *Phys Rev B* **93**, 165403 (2016).

43. Miranda VG, Dias da Silva LGGV, Lewenkopf CH. Coulomb charging energy of vacancy-induced states in graphene. *Phys Rev B* **94**, 075114 (2016).

44. Mitchell AK, Fritz L. Kondo effect with diverging hybridization: Possible realization in graphene with vacancies. *Phys Rev B* **88**, 075104 (2013).

45. Huertas-Hernando D, Guinea F, Brataas A. Spin-orbit coupling in curved graphene, fullerenes, nanotubes, and nanotube caps. *Phys Rev B* **74**, 155426 (2006).

46. Ren J*, et al.* Kondo Effect of Cobalt Adatoms on a Graphene Monolayer Controlled by Substrate-Induced Ripples. *Nano Letters* **14**, 4011-4015 (2014).

47. Balakrishnan J, Kok Wai Koon G, Jaiswal M, Castro Neto AH, Ozyilmaz B. Colossal enhancement of spin-orbit coupling in weakly hydrogenated graphene. *Nat Phys* **9**, 284-287 (2013).

48. Li G, Luican-Mayer A, Abanin D, Levitov L, Andrei EY. Evolution of Landau levels into edge states in graphene. *Nat Commun* **4**, 1744 (2013).

49. Lu C-P*, et al.* Local, global, and nonlinear screening in twisted double-layer graphene. *PNAS* **113**, 113, 6623-6628 (2016).

50. Luican-Mayer A*, et al.* Screening Charged Impurities and Lifting the Orbital Degeneracy in Graphene by Populating Landau Levels. *Phys Rev Lett* **112**, 036804 (2014).

## Acknowledgments

We acknowledge support from DOE-FG02-99ER45742 (E.Y.A. and J.M.), NSF DMR 1708158 (Y.J.), Ministry of Science and Technology and also Academia Sinica of Taiwan (G.Y.G. and P.W.L.), Deutsche Forschungsgemeinschaft via project AN-275/8-1 (D.M. and F.B.A.), Key Research Program of the Chinese Academy of Sciences XDPB08-1 (J.M.). We thank Natan Andrei, Bruno Uchoa and Mohammad Sherafati for useful discussions.

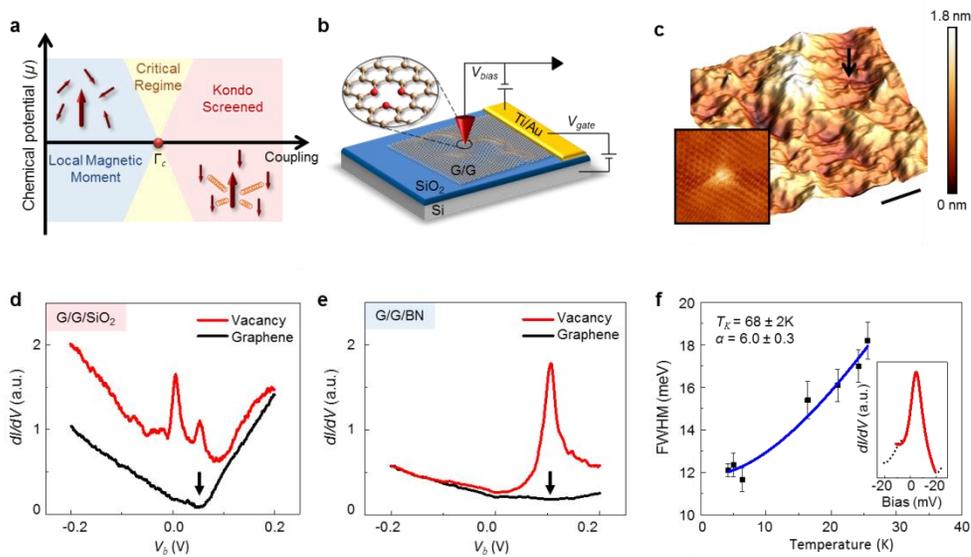

**Fig. 1 Kondo peak at a single-atom vacancy in graphene.** (**a**) Schematic phase diagram of the pseudo-gap Kondo effect. The critical regime (yellow) separates the Local-magnetic-moment phase from the Kondo-screened phase. Arrows represent the ground state of the system with the large arrow corresponding to the local spin and the smaller ones representing the spins of electrons in the conduction band. (**b**) Schematics of the experimental setup. (**c**) STM topography of a double layer graphene on $SiO_2$ (G/G/$SiO_2$). The arrow indicates an isolated vacancy ($V_b$ = -300mV, $I$ = 20pA, $V_g$ = 50V). The scale bar is 20 nm. Inset: atomic resolution topography of a single atom vacancy shows the distinctive triangular structure (4nm x 4nm), $V_b$ = -200mV, $I$ = 20pA, $V_g$ = -27V. (**d**) $dI/dV$ spectra at the center of a single atom vacancy (upper red curve) and on pristine graphene far from the vacancy (lower black curve). The curves are vertically displaced for clarity ($V_b$ = -200mV, $I$ = 20pA, $V_g$ = 0V). The arrow labels the Dirac point. (**e**) Same as (d) but for a vacancy in a G/G/BN sample ($V_b$ = -200mV, $I$ = 20pA, $V_g$ = -30V). (**f**) Evolution of the measured full width at half maximum (FWHM) with temperature (black data points) shown together with the fit (blue solid line) discussed in the text. Error bars represent the linewidths uncertainty obtained from fitting the Kondo peak to a Fano lineshape. Inset: Zoom into the Kondo peak (black dotted line) together with the Fano lineshape fit (red solid line).

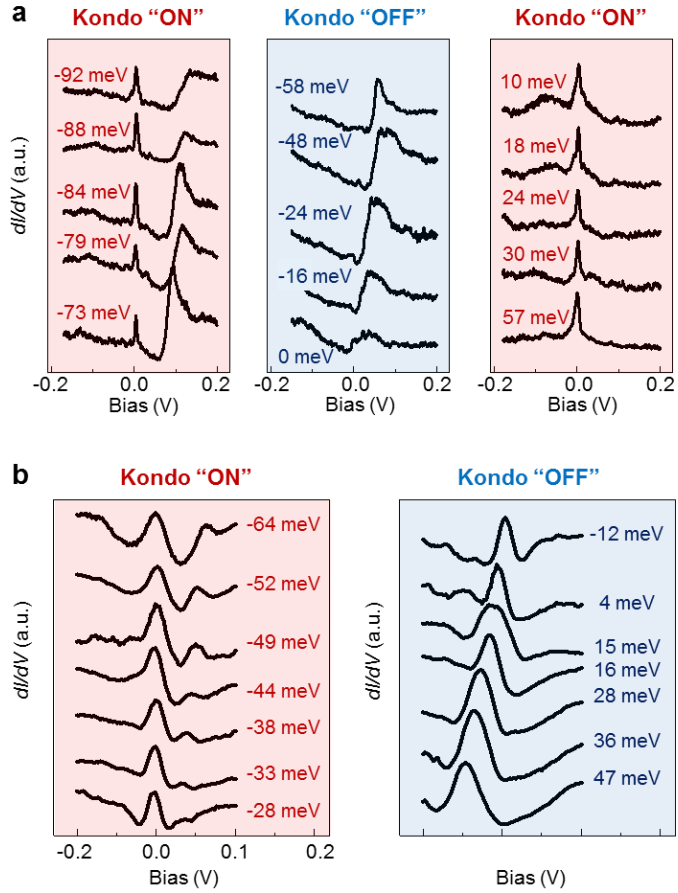

**Fig. 2 Evolution of Kondo screening with chemical potential.** (**a**) $dI/dV$ curves for a subcritical Kondo vacancy (type I in text) with reduced coupling strength $\Gamma_0/\Gamma_C = 0.90$ at the indicated values of chemical potential. Red (blue) shade indicates the presence (absence) of the Kondo peak ($V_b = -200$mV, $I = 20$pA). The chemical potential is tuned by the backgate voltage[28]. (**b**) $dI/dV$ curves for a supercritical Kondo vacancy (type II in text) with $\Gamma_0/\Gamma_C = 1.83$.

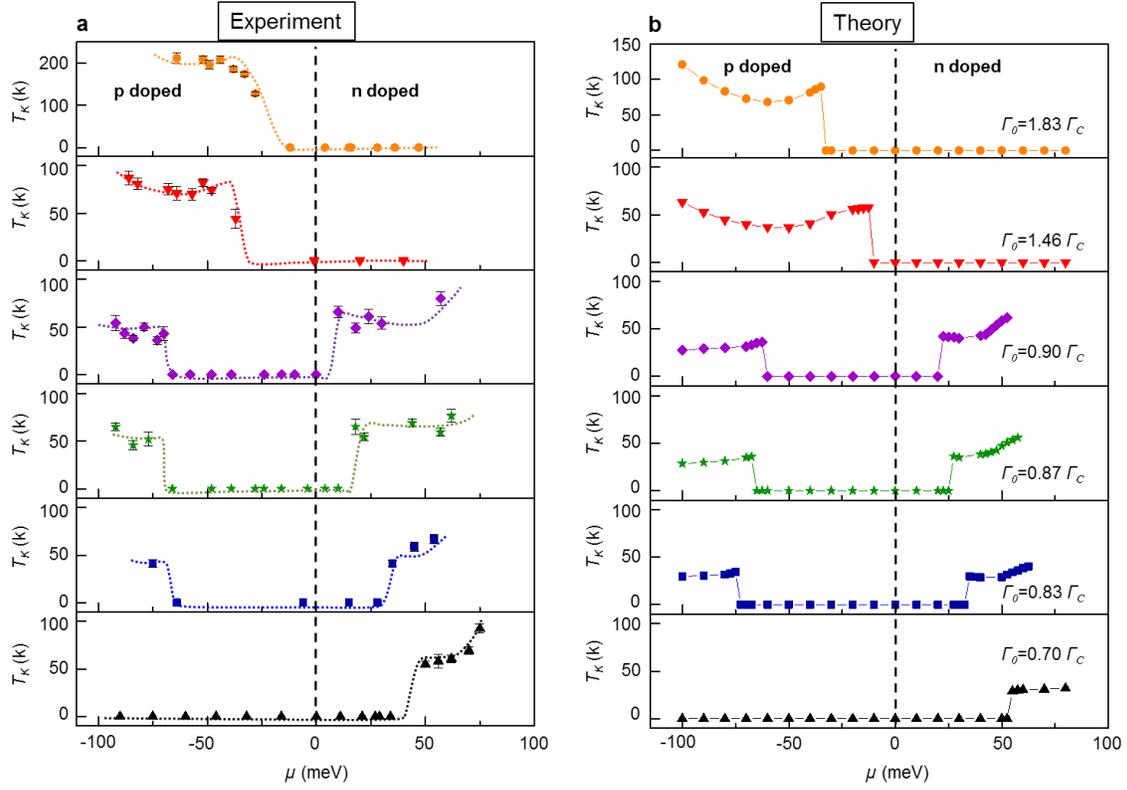

**Fig. 3 Chemical-potential dependence of the Kondo temperature**. (**a**) Chemical potential dependence of $T_K$ obtained from the Fano lineshape fit of the Kondo peak. In the regions where the peak is absent we designated $T_K = 0$. (**b**) NRG result for the vacancies in panel (a). $T_K$ is estimated by fitting the numerically simulated Kondo peak (Supplementary Note 3).

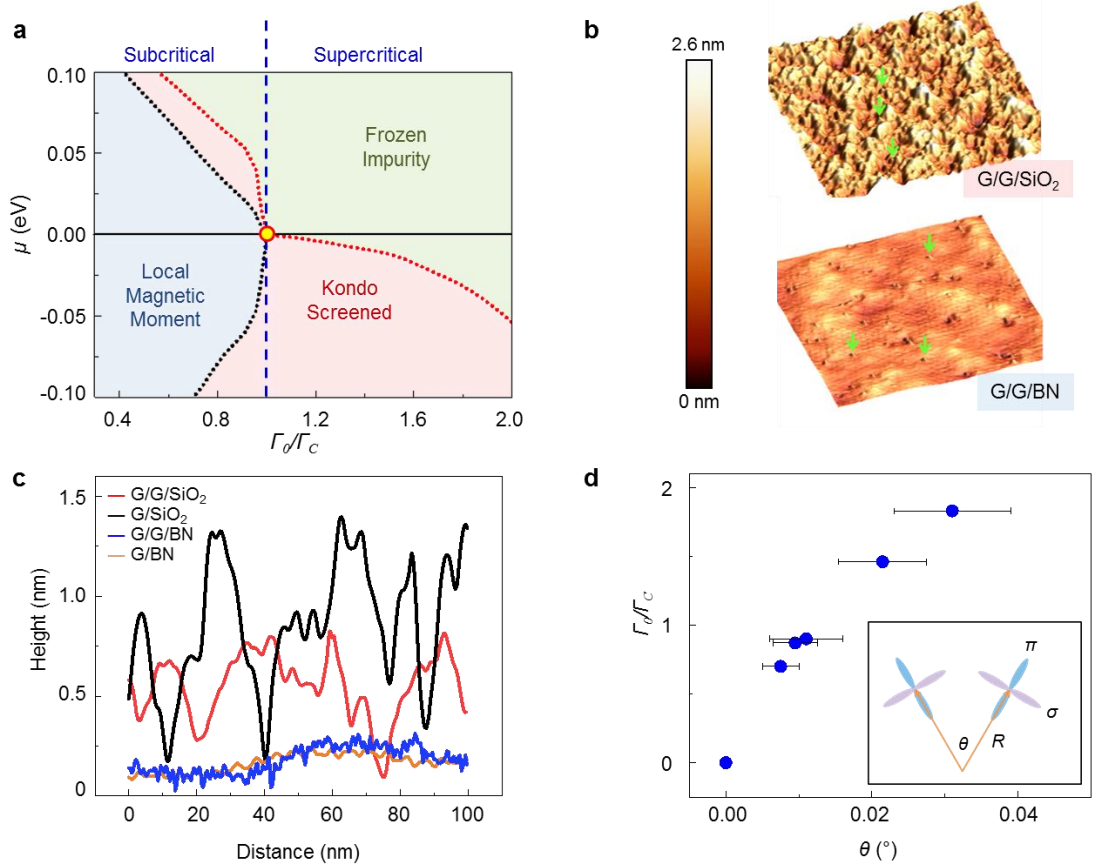

**Fig. 4 Quantum phase transition and Kondo screening.** (**a**) $T_K$–$\Gamma_0$ phase diagram at 4.2K. The critical coupling $\Gamma_C$ (circle at $\Gamma_0/\Gamma_C$ =1.0) designates the boundary between Frozen-Impurity and the Local-Magnetic-Moment phases at $\mu$=0. Dashed lines represent boundaries between the phases (Supplementary Note 8). (**b**) STM topography for the G/G/SiO$_2$ (top) and G/G/BN (bottom) samples with the same scale bar ($V_b$ = -300mV, $I$ = 20pA). The arrows point to the vacancies. (**c**) Typical line profile of the STM topographies of graphene on different substrates with the same scanning parameters as in (b). (**d**) The evolution of the hybridization strength with the curvature. Error bars represent the uncertainty in obtaining the angle between the σ-orbital and the local graphene plane orientation from the local topography measurements. Inset: sketch of the curvature effect on the orbital hybridization.

Supplementary Information

# Inducing Kondo Screening of Point Defects in Graphene with Gating and Local Curvature

Jiang *et al.*

## Supplementary Note 1: Electronically decoupled top layer graphene.

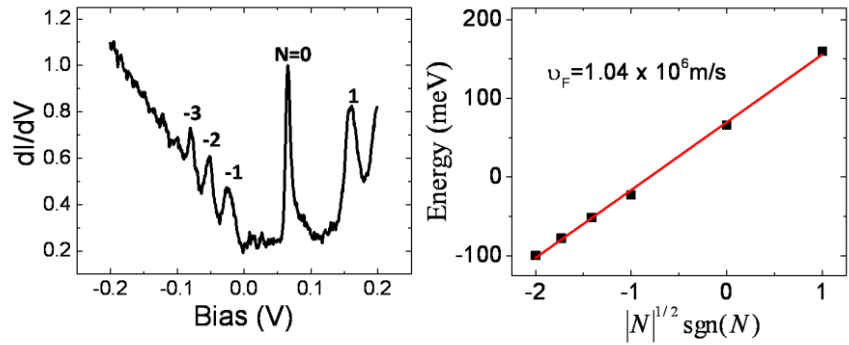

**Supplementary Figure 1**. (Left) Landau level (LL) spectroscopy of graphene in magnetic field of B = 6T. The numbers label the LL index. (Right) LL peak sequence and its fit to $E_N = E_D \pm \frac{\hbar v_F}{l_B}\sqrt{2|N|}$, $N = 0, \pm 1, ..$ is used to extract the Fermi velocity.

The two graphene layers were stacked with a large twist angle between them to ensure electronic decoupling [1, 2]. This absence of moiré patterns for these samples is consistent with their decoupling. The single layer nature of the top layer is further revealed by the characteristic Landau level (LL) sequence of peaks, $E_N = E_D \pm \frac{\hbar v_F}{l_B}\sqrt{2|N|}$, $N = 0, \pm 1, ..$ which appears in the presence of a magnetic field. Here $E_D$ is the Dirac point (DP) energy measured relative to $E_F$, $v_F$ is the Fermi velocity and $l_B$ the magnetic length. Supplementary Figure 1 shows the LL spectrum on the graphene surface far from any vacancy from which we extract the Fermi velocity $v_F = 1.04 \times 10^6$ m s$^{-1}$ by fitting to the LL sequence. The specific LLs sequence as the fingerprint of massless Dirac fermions is the direct proof of that the two graphene layer are electronically decoupled.

## Supplementary Note 2: Criteria to identify the intrinsic single vacancy

As discussed in the main text, several criteria were used to identify the intrinsic single atom vacancies. Firstly, theoretical and experimental work has shown that single atom vacancies display a characteristic triangular structure arising from the electronic reconstruction [3, 4]. This triangular fingerprint is used to distinguish between bare vacancies and vacancies that are passivated by adsorbed adatoms such as hydrogen or nitrogen whose topography lacks this feature [5, 6]. Secondly, the zero mode peak at the Dirac point allows to distinguish between single-atom vacancies and other types of un-passivated defects, such as Stone-wales reconstruction or di-vacancies, which lack both the triangular electronic reconstruction feature and the zero mode in the electronic states [7, 8]. Thirdly, the evolution of the local Dirac point with the distance from the single vacancy center is examined by the dI/dV curves. If a chemical bond forms between the dangling bond and an ad-atom, the electron affinity difference will cause a charge transfer at the vacancy site. This shifts the local Dirac point and gives rise to a clear shift of the Dirac across the vacancy site [9]. Combining all these criteria together ensures the correct identification of single atom vacancies.

## Supplementary Note 3: Fano fitting to the Kondo peak

The Kondo temperature is obtained by fitting the *dI/dV* curve to the Fano lineshape $\frac{dI}{dV} \propto \frac{(\varepsilon+q)^2}{1+\varepsilon^2} + A$, where *A* is the background tunneling signal and *q* is the Fano asymmetry factor given by $q \propto t_2/t_1$ ($t_1$ and $t_2$ are the matrix elements for electron tunneling into the continuum of the bulk states and the discrete Kondo resonance, respectively)[10]. Here $\varepsilon = \frac{eV-\varepsilon_0}{\Gamma/2}$ is the normalized energy ($\varepsilon_0$ is the position of the resonance and $\Gamma$ is the full width of half maximum (FWHM) of the Kondo peak, which is related to Kondo temperature, $T_K$, as $k_B T_K = \Gamma/2$). Due to full self-energy of the resonant level interactions and coupling to the conduction electrons, $\varepsilon_0$ may slightly shift away from $E_F$[11].

## Supplementary Note 4: Temperature dependence of the Kondo peak

The temperature dependence of the Kondo peaks is analyzed by raising the temperature from 4.2K to 25K, Supplementary Figure 2.

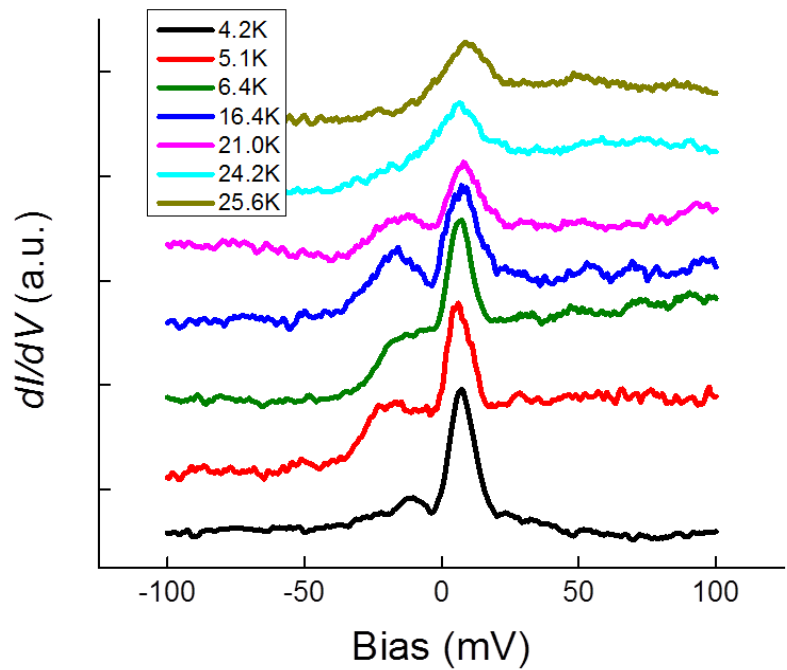

**Supplementary Figure 2** Temperature dependence of the Kondo peaks. $V_b = -200$mV, $I = 20$pA, $V_g = 50$V.

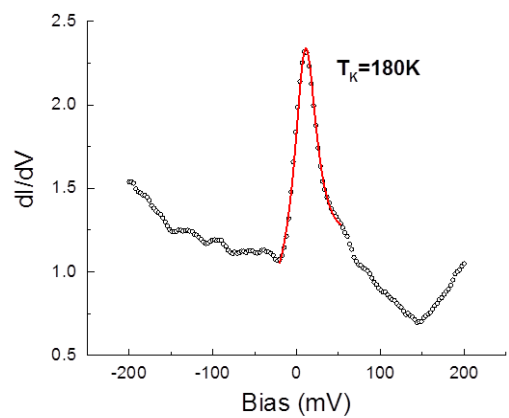

**Supplementary Figure 3** *dI/dV* curve together with Fano-fit (red) for a vacancy in G/SiO$_2$, $V_b$=-300mV, $I = 20$pA.

# Supplementary Note 5: Substrate effect on the Kondo coupling

As discussed in the main text and consistent with the theoretical predictions, Kondo screening is observed for the rougher G/SiO$_2$ and G/G/SiO$_2$ samples while it is absent on the much smoother G/BN and G/G/BN. Interestingly, most of the earlier work focused on the in-plane Jahn-Teller distortion but not on the out-of plane distortion. This is the first work to show the crucial role of the local curvature. Supplementary Figure 3 shows the Kondo peak for a vacancy in the G/SiO$_2$ sample where the local curvature is largest. T$_K$ here is much higher than that on G/G/SiO$_2$ consistent with the rougher surface.

# Supplementary Note 6: Pseudogap Anderson impurity model

As discussed in the main text, the interaction of a magnetic moment with the pseudogap conduction band electrons in graphene is captured by the single-channel asymmetric Anderson impurity model (AIM) [12] with the characteristic parameters $\varepsilon_d$, U, and Γ$_0$ corresponding to the energy of the bare impurity state, the onsite Coulomb repulsion and the coupling strength between respectively.

The Hamiltonian for the AIM can be written as[13]:

$$H = \sum_\sigma \int_{-D}^{D} d\omega \frac{(\omega+\mu)}{D} c_\sigma^\dagger(\omega) c_\sigma(\omega) + \sum_\sigma (\varepsilon_d - \mu) f_\sigma^\dagger f_\sigma + U_{eff}(\mu) f_\uparrow^\dagger f_\uparrow f_\downarrow^\dagger f_\downarrow +$$

$$\sum_\sigma \int_{-D}^{D} d\omega \sqrt{\frac{\Gamma(\omega)}{\pi D}} [c_\sigma^\dagger(\omega) f_\sigma + f_\sigma^\dagger c_\sigma(\omega)], \qquad (1)$$

where spin index $\sigma = \uparrow, \downarrow$; $D$ is the conduction bandwidth; $c_\sigma^\dagger(\omega)[c_\sigma(\omega)]$ is the creation (annihilation) operator for an electron in the conduction state with energy $\omega$; $\mu$ is the chemical potential; $\varepsilon_d$ is the energy of the impurity level; $U$ is the Coulomb interaction between the electrons on the impurity; $\Gamma(\omega)$ is the coupling function; $f_\uparrow^\dagger$ ($f_\downarrow^\dagger$) and $f_\uparrow$ ($f_\downarrow$) are creation and annihilation operators for an electron in the $\uparrow$ ($\downarrow$) impurity state.

Since the zero-mode (ZM) is near the Dirac point [see Fig. 1d and Fig. 1e in the main text] it does not contribute to the Kondo critical behavior in the p-doped regime ($\mu < 0$) as long as the ZM is unoccupied. In the n-doped regime ($\mu > 0$) however when the ZM becomes occupied, the energy of the doubly occupied σ-orbital could rise above the onsite Coulomb interaction

$U$ of the σ-orbital due to the additional Coulomb repulsion from the ZM. This increases the effective Coulomb interaction of the σ-orbital state. Therefore, we take this into account by adopting the following effective Coulomb interaction between the electrons on the impurity:

$$U_{eff}(\mu) = \begin{cases} U & \mu \leq 0 \\ U + \min(U_{\pi d}, \alpha\mu) & \mu > 0 \end{cases} \quad (2)$$

where $U_{\pi d}$ is the Coulomb interaction between the ZM and σ-orbital; $\alpha$ is a positive constant.

The coupling function can be written as [13]:

$$\Gamma(\omega) = \frac{\Omega_0 V^2 |\omega+\mu|}{2\hbar^2 v_F^2}\left[2 - J_0\left(\frac{2}{3}\frac{|\omega+\mu|}{t}\right)\right], \quad (3)$$

where $\Omega_0, V, t,$ and $v_F$ are the unit cell area, the hybridization strength, hopping energy, and Fermi velocity, respectively. We can expand the zeroth Bessel function $J_0$ at $\omega = 0$, and $\Gamma(\omega)$ can be approximated as:

$$\Gamma(\omega) = \frac{\Omega_0 V^2 |\omega+\mu|}{2\hbar^2 v_F^2}\left[1 + \frac{4}{27}\left(\frac{|\omega+\mu|}{t}\right)^2\right]. \quad (4)$$

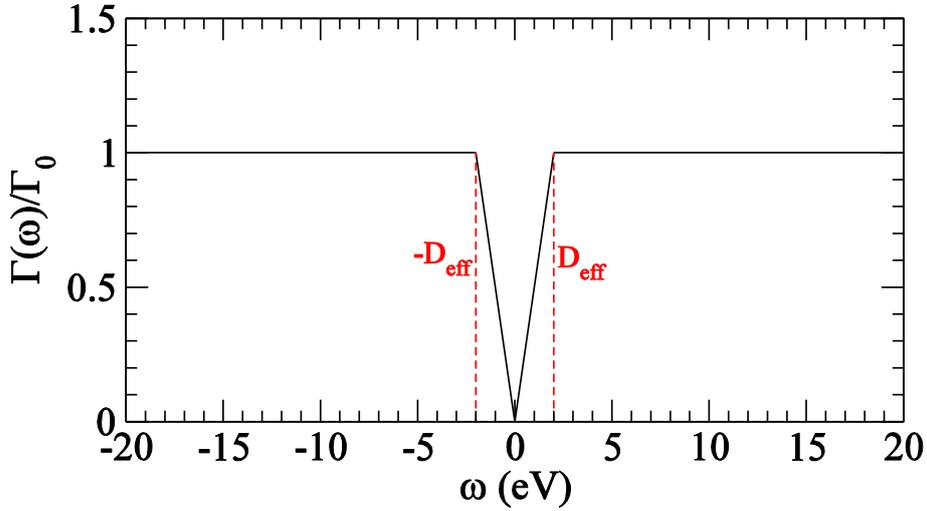

**Supplementary Figure 4.** The coupling function $\Gamma(\omega)$ adopted in this work. The total bandwidth $D$ = 20 eV and effective bandwidth $D_{eff}$ = 2 eV are taken from the graphene band structure from ab initio density functional calculations.

Earlier, we showed that the second term hardly affects the calculated spectral function[14], and thus we approximate the $\Gamma(\omega)$ as a linear function of ω. In this work, we use the mixed form in which $\Gamma(\omega)$ is proportional to $|\omega + \mu|$ only within the effective bandwidth $D_{eff}$, and is constant for the rest of the bandwidth (Supplementary Figure 4). Mathematically, $\Gamma(\omega)$ can be written as:

$$\Gamma(\omega) = \begin{cases} \Gamma_0 \frac{|\omega+\mu|}{D_{eff}} & \frac{|\omega+\mu|}{D_{eff}} \leq 1 \\ \Gamma_0 & \frac{|\omega+\mu|}{D_{eff}} > 1, \frac{|\omega|}{D_{eff}} \leq \frac{D}{D_{eff}} \\ 0 & \frac{|\omega|}{D_{eff}} > \frac{D}{D_{eff}} \end{cases} \quad (5)$$

where $\Gamma_0 = \frac{\Omega_0 V^2 D_{eff}}{2\hbar^2 v_F^2}$ represents the coupling strength discussed in the main text.

## Supplementary Note 7: Numerical renormalization group calculations

We exploit the powerful numerical renormalization group (NRG) method to solve the Anderson impurity model[15]. In the present calculations, we use the discretization parameter $\Lambda$ = 1.8 and keep 1200 states per NRG iteration so that the obtained spectral functions converge within 0.1 %. We set the total bandwidth $D = 20$ eV and $D_{eff} = 2$ eV [14] (Supplementary Figure 4). Note that the numerical results do not depend on $D$ and $D_{eff}$ as long as they are larger than 8 eV and 1 eV, respectively. Typically, the onsite Coulomb interaction varies from about 1 to 10 eV. Our experimental results indicate that the singly occupied impurity state ($\varepsilon_d$) is well below the Fermi level and only the doubly occupied impurity state ($U + \varepsilon_d$) is relevant here (i.e., $|U + \varepsilon_d| < |\varepsilon_d|$). Starting with the initial value $U = 2$ eV for the Coulomb interaction strength and $\varepsilon_d = -1.5$ eV for the bare σ-orbital energy [14, 16], we fit several theoretical spectral functions for slightly varied $U$ and $\varepsilon_d$ values to a typical experimental $dI/dV$ curve, resulting in best fits for $\varepsilon_d = -1.6$ eV and $U = 2$ eV. With this set of the model parameters, we performed the NRG calculations of the spectral function over a range of $\Gamma_0$ values in order to determine the critical value $\Gamma_C$ which separates the local-moment (LM) and frozen-impurity (FI) regimes at $\mu = 0$. The two regimes are identified by the value of the impurity entropy, S(LM) = ln2 and S(FI)=0 for the LM and FI regimes respectively. We found $\Gamma_C$ =1.15 eV from the results of these NRG calculations. Finally, by matching the spectral functions to the experimental spectra at different values of the chemical potential (Supplementary Figure 5), we obtain $\alpha=3.5$ and $U_{\pi d} = 0.2$ eV.

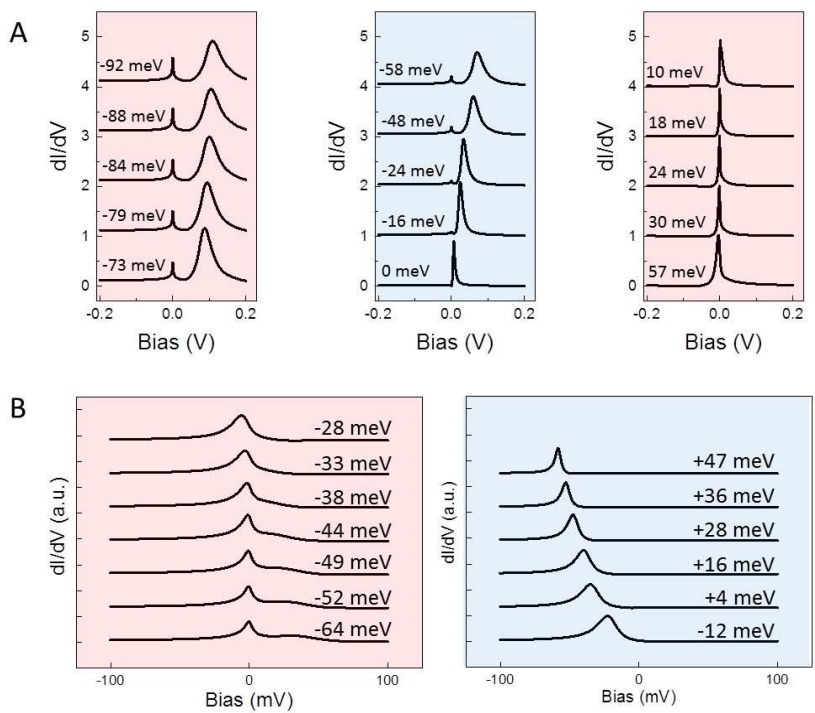

**Supplementary Figure 5.** Simulated gate-dependent *dI/dV* spectra for (A) $\Gamma_0 = 0.9\Gamma_C$ and (B) $\Gamma_0 = 1.83\Gamma_C$ corresponding to the experimental spectra in Fig. 2 of the main text.

We estimated the Kondo temperatures by fitting the spectral functions calculated at the experimental temperature T = 4.2 K to the Fano lineshape. Fig. 3a shows the Kondo temperature versus the chemical potential for six different vacancies together with the simulated results. For $\Gamma_0 < \Gamma_C$, Kondo screening occurs in both p-doped and n-doped sectors. Supplementary Figure 5A shows the simulated gate-dependent *dI/dV* spectra for (A) $\Gamma_0 = 0.90\Gamma_C$ and (B) $\Gamma_0 = 1.83\Gamma_C$ corresponding to the experimental spectra in Fig. 2 of the main text. The NRG simulations of the vacancy spectra in Fig.2a (main text), correspond to $\Gamma_0 = 0.90\Gamma_C$, which places this vacancy in the subcritical regime. In Supplementary Figure 5B, the NRG simulations of the vacancy spectra in Fig. 2b (main text), correspond to $\Gamma_0 = 1.83\Gamma_C$, which places this vacancy in the supercritical regime. A comparison between the measured (Fig. 3a) and simulated (Fig. 3b) $T_K(\mu)$ curves in main text was used to obtain the $\Gamma_0/\Gamma_C$ values characterizing each vacancy.

## Supplementary Note 8: NRG Phase diagram

The Anderson impurity model has a rich phase diagram where both spin and charge fluctuations play a role. In the case of graphene, where the particle-hole symmetry is broken by the finite value of the next nearest neighbor hopping term or for $U \neq 2|\varepsilon|$, the interaction of the vacancy spin with the linear DOS is captured by the asymmetric Anderson impurity model [17, 18, 19]. The NRG phase diagram for this model is controlled by the valence fluctuation critical point, $\Gamma_c$. At charge neutrality, μ=0, $\Gamma_C$ separates the NRG flow into two sectors: (i) supercritical, $\Gamma_0 > \Gamma_C$, which flows to the ASC fixed point where charge fluctuations give rise to a frozen impurity (FI) correlated ground state. (ii) subcritical, $\Gamma_0 < \Gamma_C$, which flows to the local LM fixed point where the impurity moment is unscreened.

In the supercritical regime, the NRG analysis reveals that the FI is a correlated ground state in which coupling between the impurity and the conduction band electrons is mediated by a cloud of charge fluctuations, resulting in a doubly occupied singlet where spin fluctuations are frozen. In the subcritical regime it was shown [18] that the impurity moment is Kondo screened for any finite value of the chemical potential, $\mu \neq 0$, but at the same time $T_K$ is exponentially suppressed: $\ln T_K \propto -1/|\mu|$. As a result, at sufficiently low doping the value of $T_K$ must fall below any experimentally accessible temperature, so that for all practical purposes its value can be set at 0. As shown in Figure 5 of reference [18], for the experimental parameters relevant to this work ($T = 4.2K$ and $\Gamma < 0.9\ \Gamma_c$), the crossover between screened and unscreened moments as a function of doping is very sharp, which allows us to label the different regimes as distinct phases. The phase diagram presented in Fig. 4A of the manuscript classifies the various regimes at the experimental temperature of 4.2 K using this criterion.

*Kondo model.* In the limit $|\varepsilon_d|, |U + \varepsilon_d| \gg \Gamma, k_B T$ the Anderson impurity model reduces to the well-known spin ½ Kondo model where charge fluctuations are quenched. In this case the single occupancy of the impurity level is overwhelmingly favored over zero or double occupancy, in effect localizing a pure-spin degree of freedom at the impurity site. As a result screening of the impurity moment is mediated by spin fluctuations only. For a linear DOS such as graphene and with a particle-hole asymmetric band, the Kondo model NRG phase diagram is characterized by a critical coupling strength that separates a local moment (LM) phase from a strongly coupled (ASC) phase where the moment is screened by interactions with the conduction band that are mediated by spin-fluctuations. The dynamics of the Kondo screening cloud is thus distinct from that in the FI where the screening is mediated by charge fluctuations.

## Supplementary References

1. Lu C-P, *et al.* Local, global, and nonlinear screening in twisted double-layer graphene. *PNAS* **113**, 113, 6623-6628 (2016).

2. Luican A, *et al.* Single-Layer Behavior and Its Breakdown in Twisted Graphene Layers. *Phys Rev Lett* **106**, 126802 (2011).

3. Amara H, Latil S, Meunier V, Lambin P, Charlier JC. Scanning tunneling microscopy fingerprints of point defects in graphene: A theoretical prediction. *Phys Rev B* **76**, 115423 (2007).

4. Banhart F, Kotakoski J, Krasheninnikov AV. Structural Defects in Graphene. *ACS Nano* **5**, 26-41 (2011).

5. Zhao L, *et al.* Visualizing Individual Nitrogen Dopants in Monolayer Graphene. *Science* **333**, 999-1003 (2011).

6. Ziatdinov M, Fujii S, Kusakabe K, Kiguchi M, Mori T, Enoki T. Direct imaging of monovacancy-hydrogen complexes in a single graphitic layer. *Phys Rev B* **89**, 155405 (2014).

7. Zaminpayma E, Razavi ME, Nayebi P. Electronic properties of graphene with single vacancy and Stone-Wales defects. *Applied Surface Science* **414**, 101-106 (2017).

8. Mao J, *et al.* Realization of a tunable artificial atom at a supercritically charged vacancy in graphene. *Nat Phys* **12**, 545-549 (2016).

9. Ma C, *et al.* Tuning the Doping Types in Graphene Sheets by N Monoelement. *Nano Letters*, (2017).

10. Fano U. Effects of Configuration Interaction on Intensities and Phase Shifts. *Physical Review* **124**, 1866-1878 (1961).

11. Madhavan V, Chen W, Jamneala T, Crommie MF, Wingreen NS. Tunneling into a Single Magnetic Atom: Spectroscopic Evidence of the Kondo Resonance. *Science* **280**, 567-569 (1998).

12. Gonzalez-Buxton C, Ingersent K. Renormalization-group study of Anderson and Kondo impurities in gapless Fermi systems. *Phys Rev B* **57**, 14254-14293 (1998).

13. Kanao T, Matsuura H, Ogata M. Theory of Defect-Induced Kondo Effect in Graphene: Numerical Renormalization Group Study. *J Phys Soc Jpn* **81**, 063709 (2012).

14. Lo PW, Guo GY, Anders FB. Gate-tunable Kondo resistivity and dephasing rate in graphene studied by numerical renormalization group calculations. *Phys Rev B* **89**, 195424 (2014).

15. Wilson KG. The renormalization group: Critical phenomena and the Kondo problem. *Rev Mod Phys* **47**, 773-840 (1975).

16. Yazyev OV, Helm L. Defect-induced magnetism in graphene. *Phys Rev B* **75**, 125408 (2007).

17. Fritz L, Vojta M. Phase transitions in the pseudogap Anderson and Kondo models: Critical dimensions, renormalization group, and local-moment criticality. *Phys Rev B* **70**, 214427 (2004).

18. Vojta M, Fritz L, Bulla R. Gate-controlled Kondo screening in graphene: Quantum criticality and electron-hole asymmetry. *Epl-Europhys Lett* **90**, 27006 (2010).

19. Fritz L, Vojta M. The physics of Kondo impurities in graphene. *Rep Prog Phys* **76**, 032501 (2013).